\documentclass[aps,amsfonts,floatfix,reprint,tightenlines,amssymb,superscriptaddress,longbibliography]{revtex4-2}  

\usepackage{booktabs} 
\usepackage{tabularx}
\usepackage{graphicx} 
\usepackage{amsmath}
\usepackage{dsfont}
\usepackage{subcaption}
\usepackage{enumerate}
\usepackage{tikz}
\usetikzlibrary{quantikz2}
\usepackage{adjustbox}
\usepackage{mathtools}

\usepackage{amsthm}

\theoremstyle{definition}

\usepackage[linktocpage=true,
  colorlinks=true, 
  pdfborder={0 0 0},
  linkcolor=blue,
  citecolor=red,
  filecolor=yellow,
  urlcolor=blue,
  bookmarks,
]{hyperref}
\usepackage[capitalize]{cleveref}

\usepackage{orcidlink}
\usepackage{braket}
\usepackage{bm}

\begin{document}

\preprint{APS/123-QED}

\title{Towards Improved Quantum Machine Learning for Molecular Force Fields}

\author{Yannick Couzini\'{e}\orcidlink{0000-0002-5408-8197}}
\email{yannick.couzinie@mailbox.org}
\affiliation{
        Department of Physics,
        The University of Tokyo
        Hongo, Bunkyo-ku,
        Tokyo, Japan.
}
\affiliation{
Quemix Inc.,
Taiyo Life Nihombashi Building,
2-11-2,
Nihombashi Chuo-ku, 
Tokyo,
Japan
}

\author{Shunsuke Daimon\orcidlink{0000-0001-6942-4571}} \affiliation{
Quantum Materials and Applications Research Center,
National Institutes for Quantum Science and Technology (QST)
2-12-1 Ookayama, Meguro-ku, Tokyo 152-8550, Japan
}

\author{Hirofumi Nishi\orcidlink{0000-0001-5155-6605}}
\email{hnishi@quemix.com}
\affiliation{
        Department of Physics,
        The University of Tokyo
        Hongo, Bunkyo-ku,
        Tokyo, Japan.
}

\affiliation{
Quemix Inc.,
Taiyo Life Nihombashi Building,
2-11-2,
Nihombashi Chuo-ku, 
Tokyo,
Japan
}

\author{Natsuki Ito}
\affiliation{
        NGK INSULATORS, LTD.,
        2-56 Suda-cho, Mizuho,
        Nagoya 467-8530,
        Japan
}

\author{Yusuke Harazono}
\affiliation{
        NGK INSULATORS, LTD.,
        2-56 Suda-cho, Mizuho,
        Nagoya 467-8530,
        Japan
}

\author{Yu-ichiro Matsushita\orcidlink{0000-0002-9254-5918}}
\affiliation{
        Department of Physics,
        The University of Tokyo
        Hongo, Bunkyo-ku,
        Tokyo, Japan.
}

\affiliation{
Quemix Inc.,
Taiyo Life Nihombashi Building,
2-11-2,
Nihombashi Chuo-ku, 
Tokyo,
Japan
}

\affiliation{
Quantum Materials and Applications Research Center,
National Institutes for Quantum Science and Technology (QST)
2-12-1 Ookayama, Meguro-ku, Tokyo 152-8550, Japan
}

\date{\today}

\begin{abstract}
    This study explores the use of equivariant quantum neural networks (QNN)
    for generating molecular force fields, focusing on the rMD17 dataset. We
    consider a QNN architecture based on previous research and point out
    shortcomings in the parametrization of the atomic environments. These shortcomings limit
    its expressivity as an interatomic potential and precludes transferability
    between molecules. We propose a revised QNN architecture that addresses
    these shortcomings. While both QNNs show promise in force prediction, with
    the revised architecture showing improved accuracy, they struggle with
    energy prediction. Further, both QNNs architectures fail to demonstrate a
    meaningful scaling law of decreasing errors with increasing training data.
    These findings highlight the challenges of scaling QNNs for complex
    molecular systems and emphasize the need for improved encoding strategies,
    regularization techniques, and hybrid quantum-classical approaches.
\end{abstract}

\maketitle
\section{Introduction}
Molecular dynamics (MD) simulations are a cornerstone of computational
chemistry and materials science, offering detailed insights into the time
evolution of atomic systems across various scales. These simulations depend
critically on accurate and efficient force calculations to guide atomic motion.
While empirical potentials, such as the
Lennard-Jones~\cite{lennard-jones1925ForcesAtomsIons} or Stillinger-Weber
potentials~\cite{stillinger1985ComputerSimulationLocal,
stillinger1986ErratumComputerSimulation}, are computationally efficient, they
lack the precision required for capturing complex phenomena such as bond
formation and breaking, polarization effects, and subtle quantum mechanical
interactions.

Machine-learned interatomic potentials
(MLIPs)~\cite{friederich2021MachinelearnedPotentialsNextgeneration,
        dusson2022AtomicClusterExpansion,
        glielmo2018EfficientNonparametricnbody,
        shapeev2016MomentTensorPotentials,
        drautz2019AtomicClusterExpansion,
vandermause2020OntheflyActiveLearning, jinnouchi2019PhaseTransitionsHybrid,
jinnouchi2019OntheflyMachineLearning, darby2022CompressingLocalAtomic}, trained
on quantum chemical data from first-principles calculation, have emerged as a
transformative alternative, offering a balance between accuracy and efficiency.
However, as MLIPs are trained on an increasingly broad range of materials,
their computational cost, both in training and evaluation, becomes a bottleneck.
This is due to the growing number of parameters required to represent the data
and the increasing dimensionality of feature
spaces~\cite{deng2023CHGNetPretrainedUniversal,
batatia2024FoundationModelAtomistic, barroso-luque2024OpenMaterials2024}.

Quantum computing, with its ability to explore exponentially large Hilbert
spaces, offers a paradigm shift in this landscape. Unlike classical processors,
which represent data in binary form, quantum computers process information
using qubits that exist in superpositions of states. This property enables data
representation and manipulation that may far outpace classical approaches,
particularly in tasks relying on large state spaces, such as quantum
chemistry~\cite{lloyd1996UniversalQuantumSimulators,abrams1997SimulationManyBodyFermi,kassal2008PolynomialtimeQuantumAlgorithm,childs2012HamiltonianSimulationUsing,nishi2025EncodedProbabilisticImaginarytime},
molecular dynamics~\cite{obrien2019CalculatingEnergyDerivatives,simon2024ImprovedPrecisionScaling,kassal2009QuantumAlgorithmMoleculara,obrien2022EfficientQuantumComputation}
and machine
learning~\cite{schuld2019QuantumMachineLearning,schuld2021SupervisedQuantumMachine}. Quantum machine learning (QML) holds promise for
accelerating the training and inference of MLIPs, potentially reducing
the number of required parameters and simplifying model
complexity~\cite{gil-fuster2024UnderstandingQuantumMachine}. Early theoretical
and experimental results suggest that, in specific scenarios such as QML on quantum data~\cite{chen2022ExponentialSeparationsLearning,
huang2022QuantumAdvantageLearning}, QML models may
offer computational advantages, though the extent of QML's computational
advantages when applied on classical data is
still a topic of debate~\cite{cerezo2022ChallengesOpportunitiesQuantum,
        kubler2021InductiveBiasQuantum,
        cotler2021RevisitingDequantizationQuantum,
        chen2021HierarchyReplicaQuantum,
        cerezo2024DoesProvableAbsencea}.

Integrating QML into MD
simulations~\cite{meyer2023ExploitingSymmetryVariational,
le2023SymmetryinvariantQuantumMachine, kiss2022quantumneural, dai2022quantumgaussian, lomonaco2024quantumextreme, guo2024benchmarkingof} could unlock new levels of performance
and scalability. Hybrid quantum-classical architectures could allocate quantum
processors to specialized tasks such as solving linear
systems~\cite{harrow2009QuantumAlgorithmLinear,gilyen2019QuantumSingularValue}, optimizing
neural networks, or encoding molecular
wave functions~\cite{nakaji2022ApproximateAmplitudeEncoding,moosa2023LineardepthQuantumCircuitsa,kosugi2024QubitEncodingMixturea}, while classical
high-performance computing systems handle trajectory integration and data
storage. MD simulations are particularly well-suited for such hybrid
architectures due to their dual reliance on accurate force calculations and
scalable trajectory computations.

To formulate a model for energy and force predictions, we consider an atomic
system consisting of $N_{\mathrm{a}}$ atoms with positions $\{\vec{r}_1, \dots,
\vec{r}_{N_{\mathrm{a}}}\}$
and species $\{s_1, \dots, s_{N_{\mathrm{a}}}\}$. For a given \emph{central atom} $i$, its
molecular environment is represented as $R_i = \{\vec{r}_{ji} : j \neq i\}$, where
$\vec{r}_{ji} = \vec{r}_j - \vec{r}_i$, with species information given by $S_i
= \{s_j : j \neq i\}$. A force prediction model must respect the rotational
symmetry of physical laws: given a rotation $A \in \mathrm{O}(3)$, the
molecular energy $E$ is invariant~\cite{dusson2022AtomicClusterExpansion}
\begin{align}
        E(AR_i, S_i)=E(R_i, S_i)
\end{align}
and forces transform equivariantly as
\begin{align} \vec{F}_j(AR_i, S_i):=
        -\nabla_{A\vec{r}_{j}} E(R_i, S_i) = -A \vec{F}_j(R_i, S_i).
\end{align} This requirement motivates the use of
geometric machine learning techniques~\cite{bronstein2021GeometricDeepLearning,
atz2021GeometricDeepLearning}, which have been extended to
QML~\cite{nguyen2024TheoryEquivariantQuantum,
ragone2023RepresentationTheoryGeometrica,schatzki2024TheoreticalGuaranteesPermutationequivariant}, and in particular also used in Refs.~\cite{meyer2023ExploitingSymmetryVariational,
le2023SymmetryinvariantQuantumMachine} which we use to account for symmetries in our models.

Beyond symmetries, a well-designed MLIP should also align with physical intuition. Machine learning models often struggle to capture singularities in the energy function at very short interatomic distances due to both the scarcity of reliable data in these regions and the difficulty of modeling a diverging function. However, they should still predict a steep energy increase as atoms approach a critical minimal distance 
$r_0$~\cite{dusson2022AtomicClusterExpansion}. Additionally, a good model should be transferable, meaning it can generalize across molecules that share the same bonding characteristics but have different structures.

The quantum neural network (QNN) from~\cite{meyer2023ExploitingSymmetryVariational, le2023SymmetryinvariantQuantumMachine}, which we use as a basis for our work, does not fully meet these expectations. We propose a revised approach that better aligns with classical MLIP principles. Our analysis shows that both the original and revised QNNs struggle with energy predictions for MD trajectories, effectively capturing only the mean of thermal fluctuations (see definition of \emph{meaningful learning} in \cref{sec:training}). However, force predictions are more reliable, with a slight improvement in our revised model. Despite this, neither version exhibits a clear scaling law where accuracy improves with increasing data, suggesting that further advancements are necessary for quantum MLIPs to be viable in practical applications.

In \cref{sec:theory}, we develop a QNN architecture designed to preserve the
necessary symmetries. We analyze the encoding scheme
of~\cite{meyer2023ExploitingSymmetryVariational,
le2023SymmetryinvariantQuantumMachine}, highlight its limitations, and propose
modifications tailored for MD simulations. \Cref{sec:model} details the
specific QNN architecture and its hyperparameters. \Cref{sec:training}
discusses training methodologies, while \cref{sec:results} presents a
comparative evaluation of both QNNs, demonstrating that while the revised model
shows improvement, it still falls short of classical benchmarks. Finally, we
summarize our findings and conclusions in \cref{sec:conclusion}.

\section{Theory}%
\label{sec:theory} We develop a QML model to predict
molecular energies and atomic forces while preserving fundamental symmetries.
We construct parametrized quantum circuits acting
on $N_{\mathrm{q}} = (N_{\mathrm{a}} - 1) \in 2\mathbb{N}$ qubits~(\cref{sub:initial_encoding}) and
$2(N_{\mathrm{a}} - 1)$ qubits~(\cref{sub:revised_encoding}), using learnable parameters
$\vec{\theta}$ to predict the molecular energy and atomic forces. We omit
$\vec{\theta}$ in the notation for clarity. The output $f(R_i, S_i, \vec{\theta})=f(R_i, S_i)$ serves as an
estimate of the total molecular energy $E(R_i, S_i)$ and thus must be invariant
under $A \in \mathrm{O}(3)$
transformations~\cite{dusson2022AtomicClusterExpansion}, i.e.\ we need
\begin{align}
        f(AR_i, S_i) = f(R_i, S_i).
\end{align}
As our datasets consist of single MD trajectories, we focus on rotational
SO(3) invariance rather than full reflection and rotational symmetry.

By ensuring invariance of our predicted energy we automatically get
equivariance of the predicted forces through
\begin{align}
        \nabla_{A\vec{r}_{j}} f(AR_i, S_i)
        = A \nabla_{\vec{r}_{j}} f(R_i, S_i).
\end{align}
To enforce the SO(3) symmetry constraints in MLIPs, there are two primary
approaches:  (1)
invariant descriptor functions $\vec{y}(R_i, S_i)$ as inputs to neural networks
or Gaussian processes~\cite{bartok2010GaussianApproximationPotentials,
bartok2013RepresentingChemicalEnvironments, li2024EnforcingExactPermutation},
and (2) equivariant architectures that transform together with input
coordinates and extracting invariant outputs at the final
stage~\cite{batzner2022E3equivariantGraphNeural,
le2022EquivariantGraphAttention, batatia2022MACEHigherOrder,
nguyen2024TheoryEquivariantQuantum, meyer2023ExploitingSymmetryVariational,
le2023SymmetryinvariantQuantumMachine, west2024ProvablyTrainableRotationally}.

In this work we adopt the equivariant approach. \Cref{sub:initial_encoding} presents the
equivariant QNN from~\cite{meyer2023ExploitingSymmetryVariational,
le2023SymmetryinvariantQuantumMachine}, followed by an analysis of its
shortcomings in \cref{sub:encoding_problems}, when used to predict molecular
energies. We conclude with a revised QNN in \cref{sub:revised_encoding} designed
to better suit quantum MLIPs.

\subsection{Equivariant quantum neural network}%
\label{sub:initial_encoding}
We present the construction of an equivariant quantum neural network given
in~\cite{meyer2023ExploitingSymmetryVariational,
le2023SymmetryinvariantQuantumMachine}. This encoding does not utilize the
species information $S_i$, enabling us to make predictions in the form $f(R_i)$ using $N_{\mathrm{q}}=(N_{\mathrm{a}}-1)\in 2\mathbb{N}$ qubits.
\subsubsection{Encoding layer}%
\label{ssub:init_encoding}

We prepare pairs of the $N_{\mathrm{q}}$ qubits in singlet states $\ket{\psi} =
\bigotimes_{i=1}^{N_{\mathrm{q}}/2} \ket{S}$ (see \cref{app:singlet}), where
\begin{align}\label{eqn:singlet}
        \ket{S} = \frac{1}{2}\left(\ket{01}-\ket{10}\right),
\end{align}
which is  invariant up to a global phase when applying the same rotation quantum
gate~\cite{meyer2023ExploitingSymmetryVariational} to both qubits (see
\cref{eqn:rot_invar} later for an equation). In what follows we omit
the explicit writing of the global phase as it does not influence the final
measurement result.

\begin{figure}
        \begin{adjustbox}{width=0.45\textwidth}
        \begin{quantikz}[transparent]
                \lstick[2]{$\ket{S}$}& \gate{\Phi^{(1)}(\vec{r}_{1i}, \alpha)} &\\
                                     & \gate{\Phi^{(2)}(\vec{r}_{2i}, \alpha)} & \\
                \setwiretype{n}&\vdots&\\
                \lstick[2]{$\ket{S}$}&\gate{\Phi^{(N_{\mathrm{q}}-1)}(\vec{r}_{(N_{\mathrm{q}}-1) i}, \alpha)}& \\
                                     & \gate{\Phi^{(N_{\mathrm{q}})}(\vec{r}_{(N_{\mathrm{q}})i}, \alpha)} & \\
        \end{quantikz}
        =\begin{quantikz}
                \lstick{$\ket{\psi}$}&\qwbundle{N_{\mathrm{q}}}&\gate{\Phi(R_i)}&
        \end{quantikz}
        \end{adjustbox}
        \caption{The encoding circuit together with the initial state
                $\ket{\psi}$ to
                produce the state $\ket{R_i}$.\label{fig:encoding}}
\end{figure}
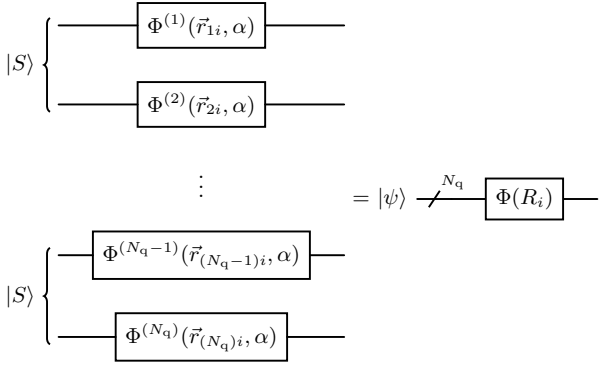

Relabel the coordinates in $R_i=\{\vec{r}_{ji}: j\in [1, \ldots, N_{\mathrm{a}}-1]\}$ by permuting the coordinates so
that $i=N_{\mathrm{a}}$.
The coordinates $\vec{r}_{ji}\in R_i$ of the $j$-th neighbour are encoded onto the
$j$-th qubit using the operator
\begin{align}\label{eqn:encoding_func}
        \Phi^{(j)}(\vec{r}_{ji}, \alpha)
        = \exp(-i\alpha \vec{r}_{ji}\cdot\vec{\sigma}^{(j)}),
\end{align}
where $\vec{\sigma}^{(j)}$ is the vector of Pauli operators $\{\mathrm{X}^{(j)},
\mathrm{Y}^{(j)}, \mathrm{Z}^{(j)}\}$ acting on the $j$-th qubit and $\alpha$ is a trainable
parameter. We denote the full encoding operation by
$\Phi(R_i)=\bigotimes_{j=1}^{N_{\mathrm{q}}}
\Phi^{(j)}(\vec{r}_{ji}, \alpha)$ and define $\ket{R_i} = \Phi(R_i)\ket{\psi}$ (see
\cref{fig:encoding}).
We now show that $\ket{R_i}$ transforms equivariantly under rotations $A \in
\mathrm{SO(3)}$, meaning that there exists a unitary representation 
$\mathcal{A}^{(j)}$ acting on the $j$-th qubit such that  
\begin{align}\label{eqn:equivariant_transform}
        \ket{AR_i} = \bigotimes_{j=1}^{N_{\mathrm{q}}} \mathcal{A}^{(j)}\ket{R_i}.
\end{align}
To construct $\mathcal{A}^{(j)}$, we express $A$ in terms of its Euler angles 
$(\nu_1, \nu_2, \nu_3)$ and decompose it into a sequence of ZXZ
rotations. The corresponding unitary operation $\mathcal{A}^{(j)}$ to $A$ for the
$j$-th qubit is then given by  
\begin{align}
        \mathcal{A}^{(j)} = \mathrm{RZ}^{(j)}(\nu_3)\mathrm{RX}^{(j)}(\nu_2)\mathrm{RZ}^{(j)}(\nu_1),
\end{align}
where we denote by RX/RY/RZ the rotation gates around the respective axes.
To verify the equivariance of the encoded state $\ket{R_i}$, consider that
$\Phi^{(j)}$ transforms under $A$ as  
\begin{align}
        \Phi^{(j)}(A\vec{r}_{ji}, \alpha) =
        \mathcal{A}^{(j)}\Phi^{(j)}(\vec{r}_{ji}, \alpha)
        \mathcal{A}^{(j)\dagger}.
\end{align}
Since singlet states remain invariant under quantum rotations~\cite{meyer2023ExploitingSymmetryVariational}, we have  
\begin{align}\label{eqn:rot_invar}
        \mathcal{A}^{(j)} \otimes \mathcal{A}^{(j+1)} \ket{S} =
        \ket{S}.
\end{align}
Thus, the full quantum state transforms equivariantly as
\cref{eqn:equivariant_transform}.

\subsubsection{Parameter layer}%
\label{ssub:init_parameter}
The parameter layer applies parametrized quantum circuits $\mathrm{L}_d(R_i)$ (see
\cref{eqn:layer}) to $\ket{R_i}$. We need to construct $\mathrm{L}_d(R_i)$ such that
equivariance is maintained, i.e.\ such that
\begin{align}\label{eqn:equivar_layer}
        \mathrm{L}_d(A R_i) \ket{AR_i} = \bigotimes_{j=1}^{N_{\mathrm{q}}} \mathcal{A}^{(j)}
        \mathrm{L}_d(R_i)\ket{R_i},
\end{align}
where $\mathcal{A}^{(j)}$ is as in \cref{ssub:init_encoding}.
For this, we use parametrized unitary
operators that commute with general rotation operators.
Following~\cite{meyer2023ExploitingSymmetryVariational}, we introduce
Heisenberg-type two-qubit parametrized gates
\begin{align}
        \mathrm{RH}^{(j, k)}(\theta) = \exp(-i \theta
        \vec{\sigma}^{(j)}\cdot\vec{\sigma}^{(k)}),\quad j,k\le N_{\mathrm{q}}.
\end{align}
The involved parameters $\theta$ are initialized at $0$, ensuring that the parametrized
gate begins as the
identity operator, an approach inspired by~\cite{grant2019InitializationStrategyAddressing} to
mitigate barren plateaus. The arrangement of RH gates varies depending on the
molecule (see \cref{sec:model}) and we write $I=\{(i_{11}, i_{12}), \ldots, (i_{M1},
i_{M2})\}$ for the set of pairs of qubits on which we apply RH gates. Thus the
parametrized layer introduces $M$ independent learnable parameters.

\begin{figure}
        \begin{quantikz}
                \lstick{$\ket{\psi}$}&\qwbundle{N_{\mathrm{q}}}&\gate{\Phi(R_i)}&
                \gate{\mathrm{L}_d(R_i)}\gategroup[1,steps=1,style={dashed,rounded
                corners, inner xsep=2pt},background,label style={label position=above right,anchor=south,yshift=-0.2cm}]{{$\times N_l$}}
                                     &\meter{\mathrm{H}^{(1,2)}}&\setwiretype{c}\rstick{$f(R_i)$}
        \end{quantikz}
        \caption{The full circuit diagram given by the function
        \cref{eqn:measurement}, the encoding circuit $\Phi(R_i)$ is given in
\cref{fig:encoding}, the layer $\mathrm{L}_d(R_i)$ in \cref{eqn:layer} and the
measurement operator H$^{(1, 2)}$ in \cref{eqn:to_measure}.%
        \label{fig:full_qnn_init}}
\end{figure}
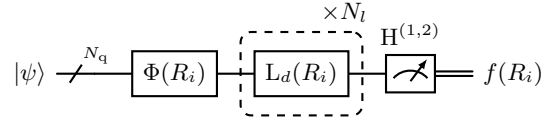
To enhance expressivity, we use data re-uploading by inserting encoding layers
$\Phi(R_i)$ after each layer of parametrized
unitaries~\cite{schuld2021EffectDataEncoding,
        perez-salinas2020DataReuploadingUniversal,
gilvidal2020InputRedundancyParameterized}. The resulting parametrized layer is
then given as
\begin{align}\label{eqn:layer}
        \mathrm{L}_d(R_i)= \Phi(R_i)\left[\prod_{k=1}^{M} \mathrm{RH}^{(i_{k1},
                i_{k2})}(\theta_{d,k}) \right] ,\; d\in \{1,
        \ldots, N_l\}
\end{align}
where $N_l$ is a hyperparameter governing the number of layers and the $R_i$
dependence comes from the re-uploading layer $\Phi(R_i)$. There are $M$ independent learnable parameters per layer, so that the parametrized layer introduces a total of
$N_l\cdot M$ independent learnable parameters. The equivariance relation
\cref{eqn:equivar_layer} follows using \cref{eqn:equivariant_transform} and
the commutation relation
\begin{align}
        \mathrm{L}_d(A R_i)\bigotimes_j \mathcal{A}^{(j)}=
        \bigotimes_j
        \mathcal{A}^{(j)}\mathrm{L}_d(R_i).
\end{align}

\subsubsection{Measurement layer}%
\label{ssub:init_measurement}
In the measurement step we need to extract SO(3)-invariant features from the
equivariant quantum state. We do this by measuring expectation values of operators that
commute with general rotations~\cite{meyer2023ExploitingSymmetryVariational}.
Specifically, we measure Heisenberg-type operators
\begin{align}\label{eqn:to_measure}
        \mathrm{H}^{(i, j)} =  \vec{\sigma}^{(i)}\cdot \vec{\sigma}^{(j)}.
\end{align}
Any measurement operator $\mathcal{O}$ that is composed of arbitrary sums of
$\mathrm{H}^{(i, j)}$ commutes with
$\bigotimes_j\mathcal{A}^{(j)}$. We choose $\mathcal{O}= \mathrm{H}^{(1, 2)}$ in
this work.  

The full QNN model is given by
\begin{align}\label{eqn:measurement}
        f(R_i)
        =\bra{R_i} \prod_{d=1}^{N_l}
        \mathrm{L}_d^{\dagger}(R_i) \mathcal{O}
        \prod_{d=1}^{N_l} \mathrm{L}_d(R_i)\ket{R_i},
\end{align}
a quantum circuit diagram is given in \cref{fig:full_qnn_init}.
It is easy to check that $f(A R_i)=f(R_i)$ using the equivariance and
commutation relations stated above. The network is trained to predict the
molecular energy, with forces obtained as their gradients. As the spectrum of
$\mathcal{O}$ is $[-3, 1]$, we need to rescale the energy labels
to fall into this range.  We rescale the energy and force labels
in the training and validation  so that the
minimum and maximum energy values fall into the range $[-1, 1]$.

\subsection{Challenges for the initial QNN}%
\label{sub:encoding_problems}

In this subsection we examine the
explicit matrix form of the single qubit encoding $\Phi(\vec{r}, \alpha)$ and
discuss its suitability as an encoding method for interatomic potentials.

Writing the single qubit state as a
two-dimensional vector, with $\ket{0}$ in the computational basis corresponding
to the top entry, we express the encoding circuit for coordinates $\vec{r}=(x,
y, z)=(r, \theta, \varphi)$ in Cartesian and spherical coordinates as
\begin{align}\label{eqn:encoding_as_matrix}
        \Phi(\vec{r}, \alpha)&=\exp(i\alpha\vec{r}\cdot \vec{\sigma})\nonumber\\
        &= \cos(\alpha r)\mathds{1} +  i \sin(\alpha r) 
        \begin{pmatrix}
                \frac{z}{r} & \frac{x-iy}{r}\\
                \frac{x+iy}{r} & \frac{z}{r}
        \end{pmatrix}\\
        &=\cos(\alpha r)\mathds{1} +  i \sin(\alpha r) 
        \begin{pmatrix}
                \cos(\theta) & \sin(\theta) e^{-i\varphi}\\
                \sin(\theta) e^{i\varphi} & -\cos(\theta)
        \end{pmatrix},
\end{align}
where for simplicity we leave out the indices of the qubit on which the
operator acts.

The special unitary group SU(2), of which $\Phi(\vec{r}, \alpha)$ is a part, is
a three-parameter Lie group, but since quantum states are defined up to a
global phase, unitary transformations acting on a single qubit effectively have
only two degrees of freedom. This corresponds to the two-sphere representation
of a qubit state on the Bloch sphere. Consequently, encoding the
three-dimensional vector $\vec{r}$ into a single-qubit unitary $\Phi(\vec{r},
\alpha)$ inherently reduces the dimensionality to two, imposing a fundamental
limit on expressibility. Notably, simply applying the same encoding circuit to
two separate qubits, as done in \cite{le2023SymmetryinvariantQuantumMachine}\cite{le2023SymmetryinvariantQuantumMachine}
and exemplified in the numerical calculations of \cref{sec:results}, does not
resolve this issue.

Additionally, as $r\to 0$, meaning the atom approaches the central atom, the
encoding circuit \cref{eqn:encoding_as_matrix} approaches the identity
operator. In contrast, classical machine-learned force fields often use
embeddings such as Gaussian
functions~\cite{schutt2017SchNetContinuousfilterConvolutional} or $0$-th order
spherical Bessel functions~\cite{gasteiger2019DirectionalMessagePassing}
\begin{align}
        \mathrm{sinc}(r) = \frac{\sin(r)}{r}.
\end{align}
These functions take their maximum value as the pairwise distances decrease,
aligning with the physical intuition that energy contributions should grow as
atoms come closer together. The current encoding, however, diminishes in
influence as distances shrink, contradicting this expectation.

The original encoding also does not incorporate atomic species information,
which contradicts the intuition that different elements contribute differently
to the energy. This omission also limits the generalization of the QNN
beyond the training molecule, as the learned parameters may not transfer
effectively to systems with different atomic compositions.

Further, note that implicit in the encoding is a periodicity in $r\rightarrow
2\pi + r$ which is inherent to quantum computation, but poses a problem to
predicting the energy of non-periodic systems such as molecules.

Finally, the original encoding presented in~\cite{le2023SymmetryinvariantQuantumMachine} had molecule-specific constructions to ensure permutation invariance between similar atoms, e.g.\ the QNN for 2H$_2$O has permutation invariance for the H atoms in a single molecule but not between molecules. We do not address this problem in this work and instead opt for a fixed arbitrary order of encoding the atoms (see \cref{app:qnn} for the explicit constructions).

\subsection{Revised equivariant quantum neural network}%
\label{sub:revised_encoding}
To address these issues, we introduce a \emph{revised QNN} strategy that better
aligns with the physical intuition about machine-learned force fields laid out
in the previous section  while maintaining rotational equivariance. This revised
construction utilizes $2N_{\mathrm{q}} = 2(N_{\mathrm{a}} - 1)$ qubits.

\begin{figure}
        \begin{quantikz}
                \lstick{$\ket{\psi}$}&\qwbundle{N_{\mathrm{q}}}&&\gate{\chi_{\mathrm{a}}(R_i, S_i)}&\\
                \lstick{$\ket{0}$}&\qwbundle{N_{\mathrm{q}}}&\gate{\mathrm{H}^{\otimes
                N_{\mathrm{q}}}}&\gate{\chi_{\mathrm{r}}(R_i, S_i)}&
        \end{quantikz}
        \caption{The modified encoding circuit to build $\ket{R_i, S_i}_{\mathrm{a}}$ (top) and $\ket{R_i, S_i}_{\mathrm{r}}$ (bottom). Note that the starting state of
                the angular register is $\ket{\psi}=\bigotimes_{i=1}^{N_{\mathrm{q}}/2} \ket{S}$.%
        \label{fig:modified_encoding}}
\end{figure}
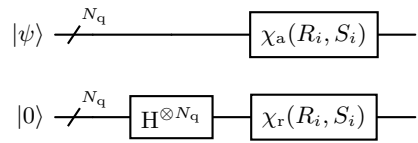
\subsubsection{Encoding layer}
Separating radial and angular degrees of freedom into distinct encodings has
been shown to enhance performance in classical machine
learning~\cite{caro2019OptimizingManybodyAtomic} and equivariant quantum neural
networks for image analysis~\cite{west2024ProvablyTrainableRotationally}. We adopt this approach
for our revised encoding strategy.

We encode the environment $R_i, S_i$ of a central atom $i$ with
position $\vec{r}_i$ and species $s_i$, using $2$ qubits per neighbor atom,
for a total of $2N_q$ qubits. Let $r_{j i} = \| \vec{r}_{ji} \|$ denote the
radial part and $\hat{r}_{ji}= \vec{r}_{j i}/ r_{j i}$ the unit vector
specifying the angular part of the relative position $\vec{r}_{j i}$.

To account for atom types in the environment and their interaction with the
central atom, we modify the original encoding~\cref{eqn:encoding_func} by
introducing learnable species-dependent parameters that depend on both the
central atom’s species $s_i$ and the surrounding atom’s species $s_j\in S_i$.
Specifically, for the angular part we define
\begin{align}
\chi_{\mathrm{a}}^{(j)} (\vec{r}_{ji}, \alpha_{s_i, s_j})
= \exp(-i \alpha_{s_i, s_j} \hat{r}_{ji}\cdot \vec{\sigma}^{(j)}).
\end{align}
This encoding is applied on the first $N_q$ qubits, prepared in pairwise
singlet states (see \cref{eqn:singlet}). The number of learnable parameters now
scales with species pairs $( s_i , s_j)$ which captures how different central
atom types interact with their neighbors, rather than being a single global
parameter.

Since the radial part $r_{j i}$ is already rotationally invariant, we adopt a
flexible species-pair-dependent encoding on the $N_a +j$-th qubit
\begin{align}
\chi_{\mathrm{r}}^{(j)}(\vec{r}_{ji}, \beta_{s_i, s_j}, \gamma_{s_i, s_j})=
\mathrm{RZ}^{(j)}(\beta_{s_i, s_j}\mathrm{sinc}(\gamma_{s_i, s_j}\cdot r_{ji})).
\end{align}
Here, $\beta_{s_i , s_j}$ and $\gamma_{s_i, s_j}$ are again indexed by both the
central and neighbouring
species, allowing for atom-type-specific radial behavior. The sinc function
ensures that short-range interactions are emphasized, aligning with classical
force-field intuition (see~\cref{sub:encoding_problems}). This operation is
applied after a Hadamard gate on each qubit initialized to
$\ket{0}$.

We call the first $N_q$ qubits the \emph{angular register}, and denote as
$\ket{R_i , S_i}_a$ the state resulting from applying the angular encoding on
the pairwise spin state $\ket{\psi}$. We call the next $N_q$ qubits the
\emph{radial register}, and denote the as  $\ket{R_i,
S_i}_r$ the state resulting from applying the radial encoding after appplying
the Hadamard gate on the radial register.

We summarize the encoding operations as:
\begin{align}
\chi_{\mathrm{a}}(R_i,
S_i)&=\bigotimes_{j=1}^{N_{\mathrm{q}}}\chi_{\mathrm{a}}^{(j)}(\vec{r}_{ji},
\alpha_{s_i, s_j})\\ \chi_{\mathrm{r}}(R_i,
S_i)&=\bigotimes_{j=N_{\mathrm{q}}+1}^{2N_{\mathrm{q}}}\chi_{\mathrm{r}}^{(j)}(\vec{r}_{ji},
\beta_{s_i, s_j}, \gamma_{s_i, s_j})
\end{align}
and the full circuit is illustrated in~\cref{fig:modified_encoding}.

\paragraph*{Remark on the central species.} In principle, the encoding must
depend on the central atom species $s_i$ to enable generalization across
chemically diverse systems. In our experiments, the systems are small enough so
that we can consider the whole system in a single calculation.  As a result,
the central atom was fixed, meaning $s_i$ remained constant and could be
omitted during training. In particular we omit it in the following sections to
keep the notation simple. However, for applications involving more complex or
heterogeneous systems, incorporating the central species  $s_i$ becomes
essential for maintaining chemical transferability.

\subsubsection{Parameter layer}
\begin{figure}
        \begin{quantikz}[transparent]
                & \gate{\mathrm{RY}(\theta_1)} &\ctrl{1}&&&&&\targ{}&\\
                & \gate{\mathrm{RY}(\theta_2)} &\targ{}& \ctrl{1}&&&&&\\
                & \gate{\mathrm{RY}(\theta_3)} &&\targ{}&\ctrl{1}&&&& \\
                \setwiretype{n}&&\vdots& &&\wire[d][1]{q} &\\
               & \gate{\mathrm{RY}(\theta_{N_{\mathrm{q}}-1})} &&&&\targ[vertical wire=q]{-1}&\ctrl{1}&& \\
                & \gate{\mathrm{RY}(\theta_{N_{\mathrm{q}}})} &&&&&\targ{}&\ctrl{-5}&\\
        \end{quantikz}
        \caption{A single RYCX$(\vec{\theta})$ block of parametrized circuits applied to the radial register. The
                 radial parameter layer $\mathrm{L}_d^{(r)}(R_i, S_i)$ consists of two such blocks, followed by a radial
                 encoding layer (\cref{fig:modified_encoding}).\label{fig:modified_layer}}
\end{figure}
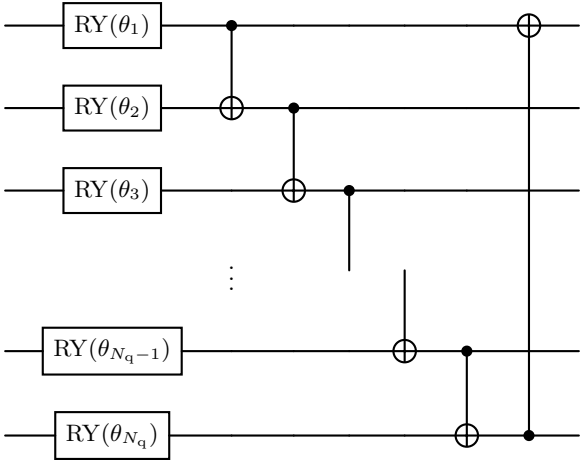
\begin{figure*}
        \centering
        \begin{quantikz}
                \lstick{$\ket{\psi}$}&\qwbundle{N_{\mathrm{q}}}&&\gate{\chi_{\mathrm{a}}(R_i,
                        S_i)}&\gate{\mathrm{L}_d^{(a)}(R_i, S_i)}\gategroup[2,steps=1,style={dashed,rounded
                corners, inner xsep=3pt},background,label style={label
                        position=above,anchor=south west,yshift=-0.2cm,
                        xshift=0.3cm}]{{$\times N_l$}}&
                \meter{\mathrm{H}^{(1, 2)}}&\setwiretype{c}\midstick{$E_a$}&\rstick[2]{$E(R_i,
                S_i)=w_a \tanh(E_a) + w_r \tanh(E_i)$}\\
                \lstick{$\ket{0}$}&\qwbundle{N_{\mathrm{q}}}&\gate{\mathrm{H}^{\otimes
                N_{\mathrm{q}}}}&\gate{\chi_{\mathrm{r}}(R_i,
                S_i)}&\gate{\mathrm{L}_d^{(r)}(R_i, S_i)}& \meter{\sum_j
                \mathrm{Z}^{(j)}}&\midstick{$E_r$}\setwiretype{c}&
        \end{quantikz}
        \caption{The modified QNN architecture.%
        \label{fig:modified_qnn}}
\end{figure*}
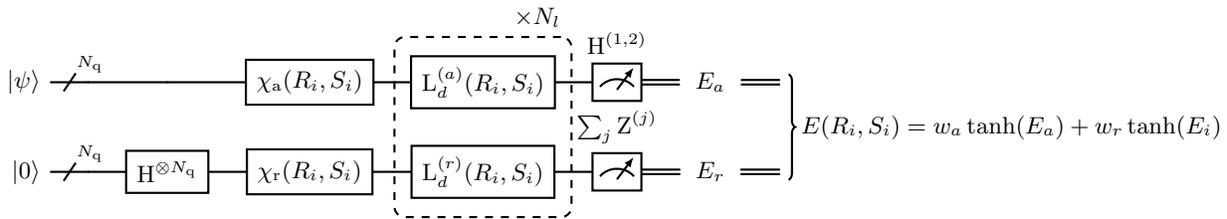
\begin{table*}[htpb] \centering
    \caption{Hyperparameters for the various QNN configurations. $N_l$ denotes
            the number of layers, $M$ is the number of independent parameters
            per layer (cf.~\cref{eqn:layer}), $M_a$ ($M_r=2N_{\mathrm{q}}$) is
            the total number of parameters per layer acting on the angular
            (radial) registers (cf.~\cref{eqn:ang_layer,eqn:rad_layer}).
            Encoding layer parameters are not included: these are $1$ for
            initial QNN and $3$ times the amount of non-central species in the
            molecule for the revised QNN, along with $2$ parameters weighting
            angular and radial contributions to the predicted energy.}
    \label{tab:hyperparameters_qnn}
    \renewcommand{\arraystretch}{1.3} 
    \begin{tabular}{l c c c c c c c c}
        \toprule
        \textbf{Molecule} & \textbf{Central Atom} &
        \multicolumn{2}{c}{\textbf{Initial QNN ($N_{\mathrm{a}}-1$)}} &
        \multicolumn{2}{c}{\textbf{Initial QNN ($2(N_{\mathrm{a}}-1)$)}} &
        \multicolumn{3}{c}{\textbf{Revised QNN}} \\
        \cmidrule(lr){3-4} \cmidrule(lr){5-6} \cmidrule(lr){7-9}
        & &  $N_l$ & $M$ & $N_l$ & $M$ & $N_l$ & $M_a$ & $M_r$ \\
        \midrule
        H$_2$O & O & 88  & 1 & 11  & 8  & 33 & 1  & 4 \\
        CH$_3$O & C & 11  & 8 & 11  & 24  & 11 & 8 & 8 \\
        CH$_3$CHO & C & 33  & 12 & 33  & 36  & 33 & 12 & 12 \\
        CH$_3$CH$_2$OH (Ethanol) & C & 11  & 24 & N/A  & N/A  & 11 & 24 & 16  \\
        \bottomrule
    \end{tabular}
\end{table*}
The angular register follows the same parameterization strategy as described in
\cref{ssub:init_parameter}, using RH gates. The angular parametrized layer
$\mathrm{L}_d^{(a)}(R_i)$ is
fully defined by the set of tuples $I=\{(i_{11}, i_{12}), \ldots, (i_{M_a1},
i_{M_a2})\}$ on which RH gates are arranged and can be written as
\begin{align}\label{eqn:ang_layer}
        \mathrm{L}^{(a)}_d(R_i, S_i)= \chi_a(R_i, S_i)\left[\prod_{k=1}^{M_a} \mathrm{RH}^{(i_{k1},
        i_{k2})}(\theta^{(a)}_{d,k}) \right],
\end{align}
where again $d\in \{1, \ldots, N_l\}$. Thus we introduce $M_a$ independent
learnable parameters per layer that act on the angular register for a total of
$M_a\cdot N_l$ parameters.
In contrast, the parameter layer acting on the radial register
incorporates entanglement through a structured layer composed of two sequential
blocks with identical circuit layouts but independent parameters. Each block,
that we denote by RYCX,
consists of a layer of parametrized RY gates applied to all qubits,
followed by a cyclic nearest-neighbour CNOT (i.e.\ controlled X) entangling layer  
\begin{align}
    \mathrm{CX}^{(N_{\mathrm{q}}+1, N_{\mathrm{q}}+2)}\mathrm{CX}^{(N_{\mathrm{q}}+2, N_{\mathrm{q}}+3)} \cdots
    \mathrm{CX}^{(2N_{\mathrm{q}}, N_{\mathrm{q}}+1)}.
\end{align}  
A diagram of a single RYCX block is shown in \cref{fig:modified_layer}. The full
layer of parametrized unitaries acting on the radial register is given by
\begin{align}\label{eqn:rad_layer}
        \mathrm{L}_d^{(r)}(R_i, S_i) = \chi_r(R_i, S_i)
        \mathrm{RYCX}(\vec{\theta}_{d, 2})\mathrm{RYCX}(\vec{\theta}_{d, 1}),
\end{align}
where $\vec{\theta}_{d, k}$, $k\in \{1, 2\}$ are $N_{\mathrm{q}}$-dimensional vectors of
independent parameters and $d\in \{1, \ldots, N_l\}$. There are thus a total of
$2N_{\mathrm{q}}$ learnable parameters per layer for a total of $2
N_{\mathrm{q}} N_l$ learnable parameters acting on the radial register. Notice in particular that there is no
gate entangling the angular and radial registers, which is necessary to
maintain SO(3) equivariance on the angular register quantum state.

\subsubsection{Measurement layer}
We measure the expectation value of the operator $\mathcal{O}^{\mathrm{(a)}}=\mathrm{H}^{(1,
2)}$ as in \cref{ssub:init_measurement} on the angular register. On the radial
register, we measure $\mathrm{Z}^{(j)}$ for all $j \in 
\{N_{\mathrm{q}}+1, \dots, 2N_{\mathrm{q}}\}$ and take the sum $\mathcal{O}^{\mathrm{(r)}}=\sum_j \mathrm{Z}^{(j)}$.
Write
\begin{align}
    E_a(R_i, S_i) &= \prescript{}{\mathrm{a}}{\bra{R_i, S_i}} \prod_d \mathrm{L}_d^{\mathrm{(a)}, \dagger}
    \mathcal{O}^{\mathrm{(a)}} \prod_d \mathrm{L}_d^{\mathrm{(a)}} \ket{R_i, S_i}_{\mathrm{a}},\\
    E_r(R_i, S_i) &= \prescript{}{\mathrm{r}}{\bra{R_i, S_i}} \prod_d \mathrm{L}_d^{\mathrm{(r)}, \dagger}
    \mathcal{O}^{\mathrm{(r)}} \prod_d \mathrm{L}_d^{\mathrm{(r)}} \ket{R_i, S_i}_{\mathrm{r}}.
\end{align}
Note that both $E_a(R_i, S_i)$ and $E_r(R_i, S_i)$ are invariant w.r.t.\ to
SO(3) rotations of the input coordinates $R_i$. The range of $E_a$ is $[-3, 1]$ and of $E_r$ is
$[-N_{\mathrm{q}}, N_{\mathrm{q}}]$. We introduce final weighting parameters
$w_a$ and $w_r$ and fit the total molecular energy as
\begin{align}
        E(R_i, S_i) = w_a \tanh(E_a) + w_r \tanh(E_r),
\end{align}
where we use $\tanh$ to introduce another layer of nonlinearity and
renormalize the outputs $E_a$ and $E_r$ to the same $[-1, 1]$ range.
A quantum circuit diagram for the calculation of $E(R_i, S_i)$ is given in \cref{fig:modified_qnn}. 
The weights are initialized as $0.001$. 

\paragraph*{Remark on expressivity and symmetry.}
In our design, the angular and radial registers are kept decoupled throughout
both encoding and parameter layers to ensure SO(3) equivariance, particularly
on the angular register state. While this separation may limit expressivity by
preventing direct quantum entanglement between the two physical degrees of
freedom, it is necessary in the current architecture to preserve symmetry.
Prior work~\cite{le2023SymmetryinvariantQuantumMachine} has explored breaking
symmetry as a way to enhance expressiveness, but we choose to stay more in line
with classical MLIP intuition, where
rotational symmetries are preserved explicitly.

Classical equivariant MLIPs and more general equivariant NN typically extract
radial and angular features separately and then combine them in a
symmetry-preserving way, often using tools such as spherical harmonics and
Clebsch-Gordan coefficients~\cite{thomas2018tensorfieldnetworksrotation,
batzner2022E3equivariantGraphNeural}. Designing analogous quantum operations
that allow for equivariant mixing of angular and radial subsystems
remains an open challenge. Such developments could significantly enhance the
expressive power of equivariant QNNs and are a promising direction for future
work.

\section{Model Setup}%
\label{sec:model}
In this section, we describe the QNN architectures used for training molecular
energies and forces on MD trajectories of individual organic molecules. Building
on the rMD17 dataset benchmark~\cite{christensen2020RoleGradientsMachinea,
batatia2024FoundationModelAtomistic}, we conduct state vector simulations of our
QNNs for molecular energy and force predictions. To keep computations feasible,
we select the smallest molecule from the dataset, CH$_3$CH$_2$OH (ethanol), and
additionally generate our own rMD17-like datasets for H$_2$O, CH$_3$O, and
CH$_3$CHO using pretrained machine learning potentials. Details of the dataset
generation and properties are provided in \cref{app:data}.

Each molecule, consisting of $N_{\mathrm{a}}$ atoms, is trained using three different QNN
architectures: two employing the initial QNN from
\cref{sub:initial_encoding} and one utilizing the revised QNN from
\cref{sub:revised_encoding}. The following subsections outline the specific
settings for each QNN setup.

\paragraph{Initial QNN} 
Molecular coordinates are given in Angstroms.  We encode the raw distances of
atoms from the central atom directly. The absolute values of the distances are
well below $\pi$ \AA, ensuring that periodicity effects do not pose an issue
(see~\cref{sub:encoding_problems}). The energy range (see \cref{tab:md_data}) falls well within the
spectrum $[-1, 3]$ of $\mathcal{O}$, allowing us to
use unnormalized energies (in eV) and forces (in eV/\AA) as training targets
\footnote{Although this approach might seem unconventional, we observed no
significant difference in performance between using normalized and unnormalized
data. However, normalized data is employed in the revised QNN.}.

To match the increased qubit count in the revised QNN, we conduct two sets
of experiments for the initial QNN: one using $N_{\mathrm{a}}-1$ qubits and another
with $2(N_{\mathrm{a}}-1)$ qubits. In the latter case, the molecular information is
redundantly encoded on both the top and bottom sets of $N_{\mathrm{a}}-1$ qubits, with an
adapted RH-gate parametrized layer establishing entanglement across all qubits.
This setup isolates the impact of a larger Hilbert space on model performance,
when comparing the initial and revised QNN.
\begin{figure*}[ht]
    \centering
    \includegraphics{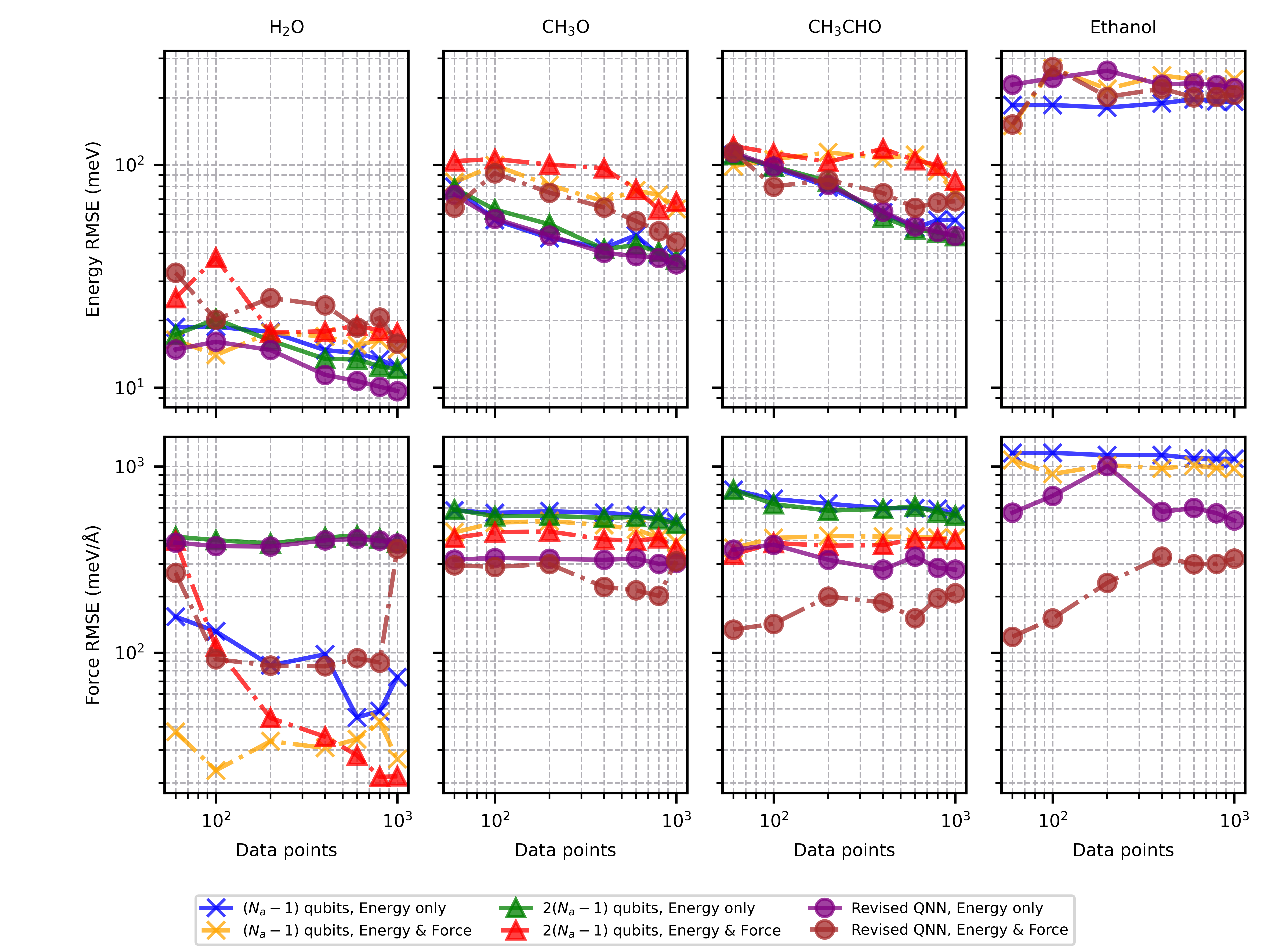}
    \caption{\label{fig:loglog} Log-log plot of the final errors
    $\sqrt{\mathcal{L}_{\mathrm{e, valid}}}$ (top) and
    $\sqrt{\mathcal{L}_{\mathrm{f, valid}}}$ (bottom) against the number
    of data points $N_{\mathrm{data}}$ on the $x$-axis for H$_2$O, CH$_3$O,
    CH$_3$CHO, and ethanol from left to right. Cold colours (solid lines) correspond to
    the \emph{Energy only} training, while warm colours (dotted lines) indicate
    the \emph{Energy \& Force} training. Crosses represent the initial QNN
    with $N_{\mathrm{a}}-1$ qubits, triangles the initial QNN with $2(N_{\mathrm{a}}-1)$ qubits,
    and circles the revised QNN.}
\end{figure*}

\paragraph{Revised QNN}
For this QNN, both input and target labels are renormalized. The input coordinates
are scaled to fit within $[-1,1]$, effectively placing the molecule inside a
cubic cell, where fractional atomic coordinates remain within $[-1, 1]$
throughout the MD trajectory. Similarly, energies (in eV) are rescaled to fall
in $[-1, 1]$, with force labels adjusted accordingly.

Based on our empirical results, given sufficient entanglement, QNN with the same
number of parameters and qubits perform similarly. Consequently, rather than
listing all 12 QNN configurations explicitly, we summarize their hyperparameters
in \cref{tab:hyperparameters_qnn}. For reproducibility we give the exact
construction details in \cref{app:qnn}. Simulations using
$2(N_{\mathrm{a}}-1)$ qubits for the initial QNN were not conducted for ethanol
due to the prohibitive computational cost of state vector simulations. In the
revised QNN, while $2(N_{\mathrm{a}}-1)$ qubits are still used, the angular and
radial registers can be simulated independently, reducing the memory requirement
to that of a $N_{\mathrm{a}}-1$ qubit circuit.

\section{Training \& Benchmarks}%
\label{sec:training}
We train our QNNs using the ADAM optimizer with a learning rate of
$0.001$ for the initial QNN and $0.005$ for the revised QNN. The
training loss consists of two components:  
\begin{itemize}
        \item $\mathcal{L}_{\mathrm{e}}$, the mean squared error (MSE) of the predicted energy.
        \item $\mathcal{L}_{\mathrm{f}}$, the MSE of the predicted forces, computed as the
            derivative of the QNN output with respect to atomic coordinates.
\end{itemize}
For the revised QNN, normalization factors are correctly accounted for by
adjusting the QNN derivatives with the cell size and normalization constant.
Training is performed with a batch size of $10$ for up to $2000$ epochs, though
we impose a 24-hour time limit. This time limit does not impose a practical
constraint on training since, given the small batch size, convergence is
typically reached within a few hundred epochs, well before the time limit is
reached. To assess the impact of including force labels in the training, we conduct separate
training runs: one minimizing only $\mathcal{L}_{\mathrm{e}}$ (\emph{Energy
only}) and another using $\mathcal{L}_{\mathrm{e}} + \mathcal{L}_{\mathrm{f}}$
(\emph{Energy \& Force}).  

We train on datasets of varying sizes, selecting $N_{\mathrm{data}} \in \{60,
100, 200, 400, 600, 800, 1000\}$ random samples from the $100,000$-step MD
trajectory. Training samples are drawn by randomly selecting indices from
the predefined $1000$ point training splits provided in the rMD17 dataset. For
validation, we evaluate errors on a fixed set of $1000$ validation points,
independent of $N_{\mathrm{data}}$, using the corresponding validation splits.
We denote the mean squared errors (MSE) on the validation energy and validation
force in the final epoch as $\mathcal{L}_{\mathrm{e, valid}}$ and
$\mathcal{L}_{\mathrm{f, valid}}$, respectively. For the \emph{Energy \& Force}
case, we include force and energy labels only for $N_{\mathrm{data}}/4$
configurations, ensuring a total of $N_{\mathrm{data}}$ labels in the training
set. From experience with classical machine learned force fields, it is expected
that the prediction accuracy on both energy and force labels is higher (and thus the loss is lower) for the
\emph{Energy \& Force} training than for the \emph{Energy only}
training~\cite{christensen2020RoleGradientsMachinea}. 
\begin{figure*}[ht]
    \centering
    \includegraphics{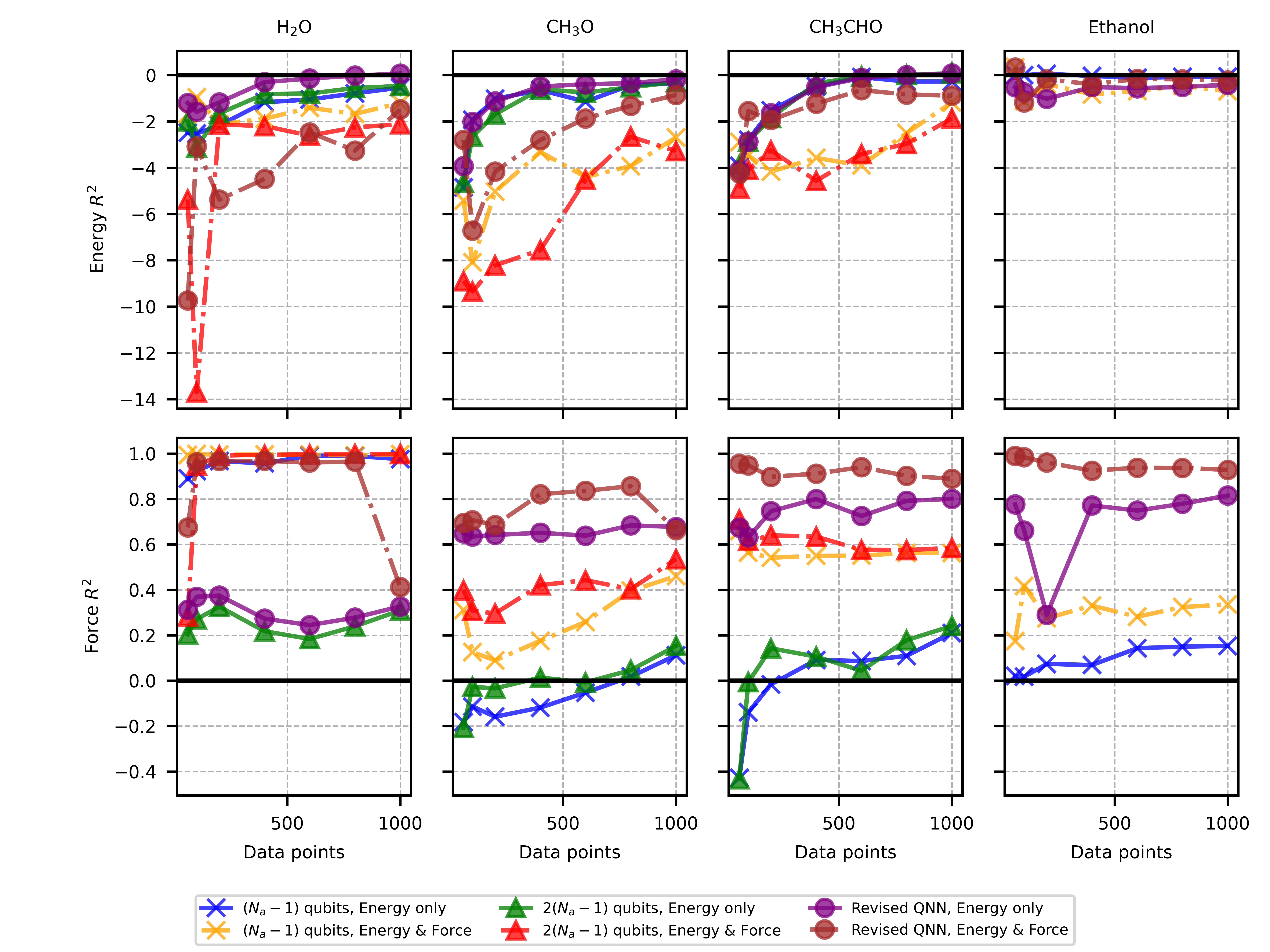}
    \caption{\label{fig:minus_std} Plot of the coefficients of determination $R^2$ given by for the energy and force data points respectively by
    $1-\mathcal{L}_{\mathrm{e, valid}}/\sigma^2_{\mathrm{E}}$
    and $1-\mathcal{L}_{\mathrm{f, valid}}/\sigma^2_{\mathrm{F}}$, where
    $\sigma_{\mathrm{E}}$ and $\sigma_{\mathrm{F}}$ denote the standard deviation
    of the energy and force data points. The black horizontal
    line marks zero, and symbols, colors, and line styles follow \cref{fig:loglog}.}
\end{figure*}

A key requirement for effective learning is that predictive accuracy should
improve systematically with an increasing number of training data, aside from
statistical fluctuations. This trend is expected when the dataset is noise-free
and the model is sufficiently
expressive~\cite{christensen2020RoleGradientsMachinea,
vapnik2013NatureStatisticalLearning, vonlilienfeld2018QuantumMachineLearning}.
Previous studies on MLIPs, particularly in the context of
rMD17, have demonstrated the scaling law  
\begin{align}
    \mathrm{Error} \propto \frac{a}{N_{\mathrm{data}}^b}.
\end{align}
The achievement of such favorable scaling is a crucial milestone in
the development of quantum MLIPs, as it
reflects the model’s ability to generalize and capture meaningful energy and
gradient representations. The authors are unaware of previous investigations
into scaling laws for quantum MLIPs.

In addition to comparing the scaling laws, we will further compare
the losses, with the standard deviations of the
energy and force labels, $\sigma_{\mathrm{E}}$ and $\sigma_{\mathrm{F}}$,
across the entire MD dataset. Since we examine a single MD trajectory,
the energy should thermodynamically fluctuate around a mean value, while the
mean of the force should be zero. We thus consider that our model exhibits
\emph{meaningful learning} only if the  coefficient of
determination $R^2$ is between $0$ and $1$, where the coefficient of determination is calculated by $R^2=1-\mathcal{L}_{X, \mathrm{valid}}/\sigma_X^2$ with $X\in \{E, F\}$ depending on whether we are looking at the energy or force errors. Otherwise, a model that
merely outputs the mean would outperform it.

\section{results \& discussion}%
\label{sec:results}
We discuss in turn the performance on energy label
prediction~\cref{sub:energy_prediction}, force label
prediction~\cref{sub:force_prediction} and discuss how well our QNNs
generalize in \cref{sub:generalize}.

\begin{table*}[ht]
    \caption{\label{tab:overfitting_table}  Differences between
    training and validation errors 
    ,
    $
        \sqrt{\mathcal{L}_{\mathrm{X, valid}}}
        -
        \sqrt{\mathcal{L}_{\mathrm{X, train}}}
    $ with $X \in \{e, f\}$, 
    for different training configurations for $N_{\mathrm{data}}=1000$, as described
    in \cref{sec:model}. 
    The energy and force errors with the smallest absolute values
    for each molecule
    and each training setting are highlighted in bold.}
    \begin{tabular*}{\textwidth}{@{\extracolsep{\fill}} l l c c}\hline\hline
System & Training setting & Energy RMSE Diff. (meV) & Force RMSE Diff. (meV/\AA) \\ \hline
\multicolumn{4}{l}{\textbf{H$_2$O}} \\ \hline
2-qubits & Energy only & 9.873 & {-12.378} \\ 
4-qubits & Energy only & 6.427 & 0.657 \\ 
Revised Encoding & Energy only & \textbf{0.461} & {-0.451} \\ 
2-qubits & Energy \& Force & 12.444 & {-2.531} \\ 
4-qubits & Energy \& Force & 15.013 & {-0.242} \\ 
Revised Encoding & Energy \& Force & 5.194 & {\textbf{-0.052}} \\ 
\hline
\multicolumn{4}{l}{\textbf{CH$_3$O}} \\ \hline
4-qubits & Energy only & 11.925 & {-0.86} \\ 
8-qubits & Energy only & 13.315 & 0.12 \\ 
Revised Encoding & Energy only & \textbf{5.173} & \textbf{0.084} \\ 
4-qubits & Energy \& Force & 30.311 & {-1.713} \\ 
8-qubits & Energy \& Force & 25.144 & {-1.023} \\ 
Revised Encoding & Energy \& Force & 7.862 & {-0.182} \\ 
\hline
\multicolumn{4}{l}{\textbf{CH$_3$CHO}} \\ \hline
6-qubits & Energy only & 17.814 & \textbf{0.285} \\ 
12-qubits & Energy only & 8.742 & {-1.886} \\ 
Revised Encoding & Energy only & \textbf{3.58} & 7.537 \\ 
6-qubits & Energy \& Force & 28.642 & 4.209 \\ 
12-qubits & Energy \& Force & 39.764 & {-2.832} \\ 
Revised Encoding & Energy \& Force & 20.252 & {-2.111} \\ 
\hline
\multicolumn{4}{l}{\textbf{Ethanol}} \\ \hline
8-qubits & Energy only & \textbf{33.17} & 2.224 \\ 
Revised Encoding & Energy only & 104.486 & {-34.811} \\ 
8-qubits & Energy \& Force & 51.774 & {-1.369} \\ 
Revised Encoding & Energy \& Force & 81.825 & {\textbf{-0.389}} \\ 
\hline
\end{tabular*}

\end{table*}

\subsection{Energy Predictions}%
\label{sub:energy_prediction}
The final validation errors are presented in \cref{fig:loglog}. Energy
prediction errors (top row) show only a slight improvement with increasing data
points, remaining within a narrow range of $50$ meV. This is in stark contrast
to classical models, which exhibit an improvement of orders of magnitude with more
training data~\cite{christensen2020RoleGradientsMachinea}. Furthermore, energy
errors increase with molecular complexity, with ethanol exhibiting the highest
errors.

Inclusion of force labels in training does not improve energy predictions.
Instead, \emph{Energy only} models (solid lines) generally achieve the best
energy accuracy. Notably, both the initial and revised encodings yield nearly
identical energy prediction performance, with no clear advantage observed for
either approach. The degradation of performance when including force labels hints at the possibility that the model is able to represent the actual energy values but not take the shape of the actual energy functional, so that the gradients do not match with the training data, prompting further research in more expressive model architectures.

\cref{fig:minus_std} provides further insight by plotting in the
top row the coefficient of determination $R^2$ of the energy data points. We see
that the $R^2$ are approaching $0$ from below with increasing $N$,
indicating that the model essentially learns the mean energy but fails to
capture variations due to thermal fluctuations. Thus, neither QNN
encoding exhibits meaningful learning of the energy data points.

This behavior can be understood by considering the nature of the
rMD17~\cite{christensen2020RoleGradientsMachinea} datasets, which are
single-temperature molecular dynamics trajectories of stable molecules. Most configurations lie
close to equilibrium, and energy variations primarily reflect vibrational
fluctuations. As a result, a model that simply predicts the average energy can
minimize the loss without learning the physical dependence of the energy on the
molecular configuration. While this reduces MSE, it leads to poor generalization and fails to
capture the shape of the potential energy surface. This is reflected in $R^2$
values approaching zero or becoming negative in the top half of \cref{fig:minus_std},
indicating that even a constant predictor may outperform the QNN. Capturing
meaningful physics requires a model that learns not only average values but
also energy gradients and their variation across configuration space.

\subsection{Force Predictions}%
\label{sub:force_prediction}
Unlike energy predictions, force predictions clearly benefit from
\emph{Energy \& Force} training. The bottom row of \cref{fig:loglog} shows that
force errors improve by approximately $20$ meV/\AA{} when force labels are
included. The revised QNN yields the lowest force errors, with the most
notable improvement observed for ethanol, where an order-of-magnitude reduction
is seen at $N_{\mathrm{data}}=60$.

However, increasing $N_{\mathrm{data}}$ does not always improve accuracy.
Unexpectedly, training with more data often leads to worse force predictions.
Despite this, the best force accuracy is achieved for H$_2$O, with an error on
the order of $1$ meV/\AA{}, approaching benchmark results for rMD17
~\cite{christensen2020RoleGradientsMachinea}. This suggests that H$_2$O, with
only three molecular degrees of freedom, may not be a sufficiently complex
benchmark for evaluating model performance.

\cref{fig:minus_std} confirms meaningful force learning,  as the coefficients of determination
(bottom row) fall between $0$ and $1$.
Interestingly, even for the \emph{Energy only} models, $R^2$ values for the forces lie between $0$ and $1$, suggesting that while the energy predictions fail to capture the correct fluctuation scale (and effectively learn only the mean), the learned energy function still contains useful local curvature information. This means the model captures aspects of the energy landscape shape, even if it does not predict absolute energy values accurately.
We see that the revised QNN consistently give similar or better $R^2$ values to the initial encoding calculations, justifying our physically motivated approach.
In particular for ethanol, the revised QNN significantly outperforms the initial
QNNs which is noteworthy, since the ethanol dataset is the only dataset based on
density-functional theory
calculations~\cite{kohn1965SelfConsistentEquationsIncluding,christensen2020RoleGradientsMachinea}
instead of being generated by pre-trained MLIPs (see \cref{app:data}).

\subsection{Overfitting and Generalization}%
\label{sub:generalize}
To investigate overfitting, we compare training and validation errors in
\cref{tab:overfitting_table} for the $N_{\mathrm{data}}=1000$ models. Large
positive discrepancies between validation and training errors indicate significant
overfitting, which prevents the model from generalizing beyond the training set.
Negative discrepancies may arise because the reported training error represents an average accumulated during training, whereas the validation error is evaluated on the final optimised parameters, which typically yield a lower error. Large negative discrepancies may therefore indicate that training was stopped before full convergence of the learnable parameters.
\begin{table*}[tpb]
    \centering
    \caption{Properties of MD data sets. DOF stands for molecular degrees of freedom.}%
    \label{tab:md_data}
    \renewcommand{\arraystretch}{1.3} 
    \begin{tabular}{l r r r r r r r}
        \toprule
        \textbf{Molecule} & \textbf{DOF} & \multicolumn{3}{c}{\textbf{Energies (eV)}} & \multicolumn{3}{c}{\textbf{Forces (eV/\AA)}} \\
        \cmidrule(lr){3-5} \cmidrule(lr){6-8}
        & & Range & Min. & Max. & Range & Min. & Max. \\
        \midrule
        H$_2$O            & \texttt{3}  & 0.069  & -14.810  & -14.741  & 4.487  & -2.214  & 2.273  \\
        CH$_3$O           & \texttt{9}  & 0.287  & -26.597  & -26.310  & 8.268  & -4.040  & 4.227  \\
        CH$_3$CHO         & \texttt{15} & 0.374  & -39.477  & -39.103  & 10.382 & -5.691  & 4.691  \\
        CH$_3$CH$_2$OH (Ethanol) & \texttt{21} & 1.539  & -4210.086 & -4208.551 & 18.303 & -9.341  & 8.962  \\
        \bottomrule
    \end{tabular}
\end{table*}

The revised QNN consistently reduces overfitting, and reduces the training and validation error gap in general, across all molecules where
meaningful learning is exhibited, suggesting that physically motivated QNNs
lead to more robust generalization. However, the remaining gap between training
and validation errors and the high losses discussed in the previous sections
indicates that
the model is still far from achieving a generalizable interatomic potential.

Note that for CH$_3$CHO and ethanol, the initial QNNs with \emph{Energy only}
training shows the lowest training-validation difference. However, these models
do not exhibit meaningful learning (see~\cref{fig:minus_std}). Among models that
do exhibit meaningful learning, the revised QNN remains the best performer.

Addressing overfitting in quantum machine learning remains a relatively
unexplored area~\cite{verdon2018UniversalTrainingAlgorithm}. It is still an open
question how much of the observed overfitting constitutes benign
overfitting~\cite{peters2023GeneralizationOverfittingQuantum} and how much
should be actively mitigated~\cite{kobayashi2022OverfittingQuantumMachine,
schuld2020CircuitcentricQuantumClassifiers}. In our case, we find no clear relation between reduced
overfitting and lower errors, suggesting that the
classical machine learning principle, where reduced overfitting leads to better
generalizations, may not hold in this context as found in previous research~\cite{peters2023GeneralizationOverfittingQuantum}.

\begin{figure}[ht]
    \centering
    \includegraphics[scale=0.5]{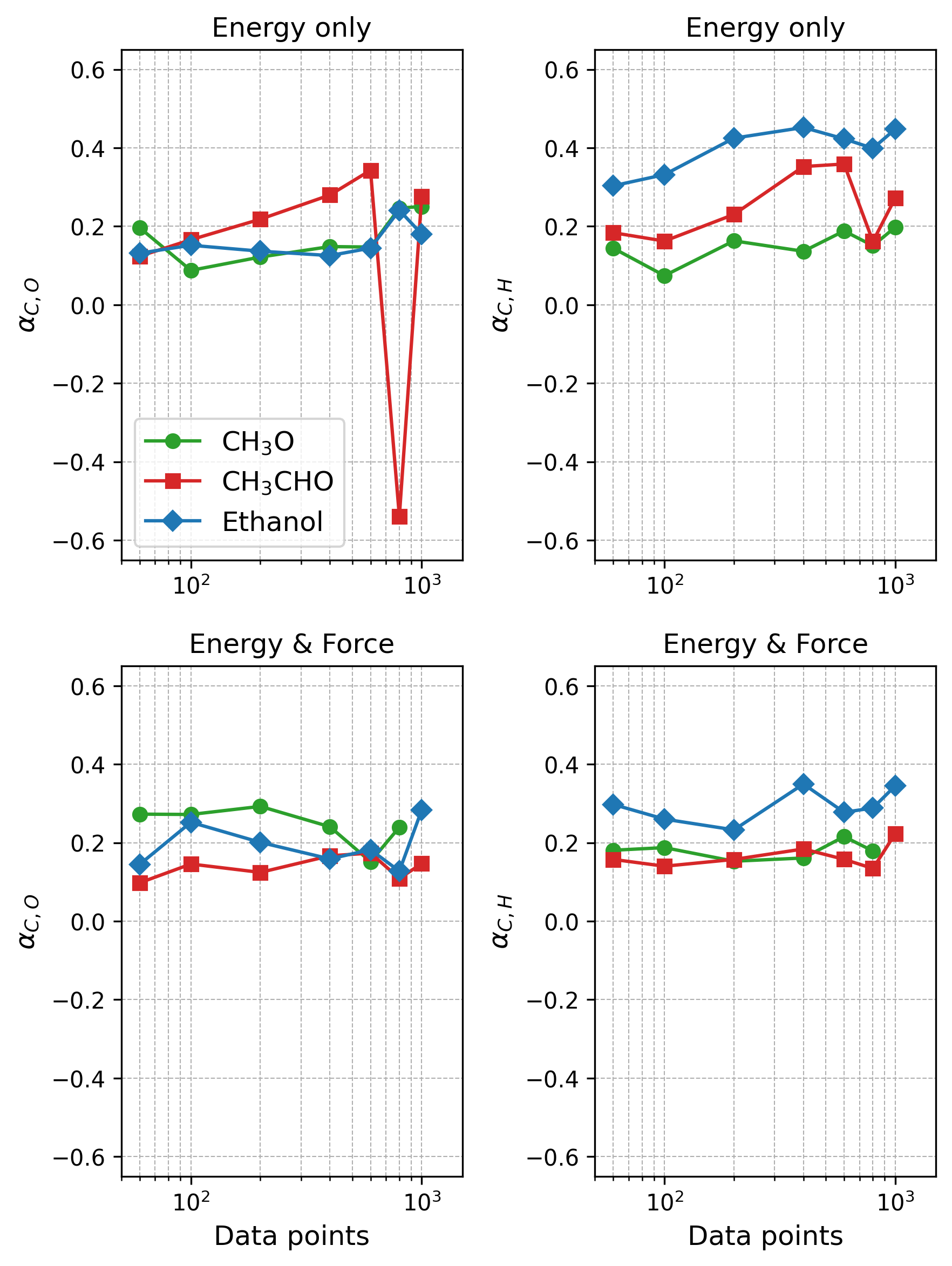} 
    \caption{Plot of the parameters $\alpha_{\mathrm{C}, \mathrm{O}}$ (left) and $\alpha_{\mathrm{C}, \mathrm{H}}$ (right) against the number of data points $N_{\mathrm{data}}$ on the $x$-axis for CH$_3$O (green circles), CH$_3$CHO (red squares), and ethanol (blue diamonds) from left to right. The upper panels corresponds to the \textit{Energy only} training, while the lower panels indicate the \textit{Energy \& Force} training.}
\label{fig:alpha_data_points}
\end{figure}

Besides generalising from training to unseen data, another relevant aspect of generalisation in our context concerns the transferability of learned parameters between different molecules that share the same atomic bond types. In \cref{fig:alpha_data_points}, we plot the learnable bond parameters $\alpha_{\mathrm{C},\mathrm{O}}$ and $\alpha_{\mathrm{C},\mathrm{H}}$ for CH$_3$O, CH$_3$CHO, and ethanol.

Despite being trained independently for each molecule, the $\alpha_{\mathrm{C},\mathrm{O}}$ parameters converge towards similar values as the number of training data points increases, suggesting a degree of consistency across molecules. In contrast, the $\alpha_{\mathrm{C},\mathrm{H}}$ parameters exhibit somewhat larger differences of about 0.1–0.2 between molecules. Given that the local electronic environments of the bonds differ considerably, further investigation is required to assess the extent of cross-molecular generalisation. Nonetheless, it is encouraging that the absolute parameter values remain relatively close, even when trained independently, indicating that the model may capture transferable bond characteristics.

\section{Conclusion}%
\label{sec:conclusion}

We investigated the application of equivariant quantum neural networks (QNNs)
to molecular potential energy and force prediction for molecular dynamics (MD)
simulations using the rMD17 dataset. While quantum machine learning (QML)
presents opportunities for advancing molecular modeling, our results highlight
key challenges that hinder current QNN approaches from achieving competitive
performance with classical methods.

A significant limitation of the initial QNN encoding was its inability to
capture three-dimensional molecular structures effectively. The encoding
inherently reduced the dimensionality of atomic environments due to the
constraints of SU(2) representation, leading to a loss of information crucial
for molecular energy prediction. Additionally, the original encoding failed to
incorporate atomic species information and diminished in influence as atoms
approached each other, contradicting physical intuition from classical force
fields.

To address these issues, we introduced a revised QNN encoding that separates
radial and angular degrees of freedom and incorporates species-dependent
learnable parameters. The angular part of the encoding applies species-specific
transformations to qubits prepared in singlet states, ensuring rotational
equivariance while maintaining flexibility. Meanwhile, the radial encoding
utilizes a sinc-based function to better capture short-range atomic
interactions, aligning with established classical embeddings. This new design
mitigates the expressivity loss of the original encoding and provides a more
physically meaningful representation of atomic interactions.

Despite these architectural improvements, our results indicate that QNNs still
struggle to achieve state-of-the-art accuracy. While
the revised encoding reduces overfitting and improves force prediction
accuracy, energy predictions remain suboptimal. The model tends to predict mean
energies rather than capture fluctuations, and the performance does not scale
significantly with increasing dataset size. This suggests that even with an
revised encoding, the QNN struggles to fully exploit additional training data
and generalize to complex molecular systems.

Moreover, we observed a trade-off between energy and force prediction accuracy:
incorporating force labels improves force predictions but degrades energy
accuracy. This imbalance suggests that QNN architectures may require additional
regularization techniques or hybrid approaches to optimize both objectives
simultaneously.

These findings underscore the current limitations of QNNs for molecular force
field generation. However, the improvements gained from our revised encoding
indicate that optimizing feature representations and network design remains a
promising research direction. Future work should explore more expressive
encoding schemes, how to more effectively include permutation
invariance of same atom kinds, ways entangle the angular and radial layers
without breaking symmetries, advanced training strategies, and hybrid quantum-classical
methods to enhance generalization and scalability.

In conclusion, while QNNs are not yet competitive with classical methods for
molecular interatomic potential generation, our study provides valuable insights
into their limitations and potential improvements. As quantum computing
advances, continued refinement of quantum machine learning techniques may
eventually lead to more accurate and scalable quantum-enhanced molecular
simulations.

\begin{table*}[htbp]
    \centering
    \caption{Qubit indices for parametrized gates. Tuples indicate RH gates,
    and quadruples represent RH4 gates, as in \cref{fig:4bitparamblock}. The
notation $\times 2$ indicates repetition of the set of indices enclosed in
brackets before it.}
    \label{tab:qnn_structures}
    \renewcommand{\arraystretch}{1.2}
    \begin{tabularx}{\textwidth}{lXX}
        \toprule
        Molecule & $N_a-1$ & $2(N_a-1)$ \\
        \midrule
        H$_2$O & {(0, 1)} & [{(0, 1, 2, 3)}] $\times 2$ \\
        CH$_3$O & [{(0, 1, 2, 3)}] $\times 2$ & {(0, 1, 2, 3), (4, 5, 6, 7), (0, 5, 6, 7), (1, 4, 6, 7), (2, 4, 5, 7), (3, 4, 5, 6)} \\
        CH$_3$CHO & [{(0, 1), (2, 3), (4, 5), (1, 2), (3, 4), (1, 5)}] $\times 2$
                  & [{(0, 1), (2, 3), (4, 5), (1, 2), (3, 4), (1, 5), (6, 7),
                  (8, 9), (10, 11), (7, 8), (9, 10), (7, 11), (0, 6), (1, 7),
          (2, 8), (3, 9), (4, 10), (5, 11)}] $\times 2$ \\
        Ethanol & {(0, 1, 2, 3), (4, 5, 6, 7), (0, 5, 6, 7), (1, 4, 6, 7), (2,
        4, 5, 7), (3, 4, 5, 6)} & N/A \\
        \bottomrule
    \end{tabularx}
\end{table*}

\begin{acknowledgments}
        This work was supported by 
        JSPS KAKENHI under Grant-in-Aid for Transformative Research Areas Grant Number JP22H05114, 
        the Center of Innovations for Sustainable Quantum AI (JST Grant Number JPMJPF2221), 
        JSPS KAKENHI under Grant-in-Aid for Early-Career Scientists Grant Number JP24K16985,
        and Japan Science and Technology Agency (JST) as part of Adopting Sustainable Partnerships for Innovative Research Ecosystem (ASPIRE), Grant Number JPMJAP24C1. 
        The computation in this work has been done using the
        supercomputer provided by Supercomputer Center at the Institute for
        Solid State Physics at the University of Tokyo (ISSPkyodo-SC-2025-Eb-0008, 2025-Ea-0013). 
\end{acknowledgments}

\section*{DATA AVAILABILITY}
The data that support the findings of this article are openly available \cite{Zenodo}.

\appendix
\section{Implementing the singlet state}
\label{app:singlet}
\begin{figure}
        \begin{quantikz}
                \lstick{$\ket{0}$} & \gate{X} & \gate{H} & \ctrl{1} & \gate{X}
                                   &\rstick[2]{$\ket{S}$}\\
                \lstick{$\ket{0}$} & \qw & \qw & \targ{} & \qw&
        \end{quantikz}
        \caption{The circuit to prepare the singlet state.\label{fig:singlet}}
\end{figure}
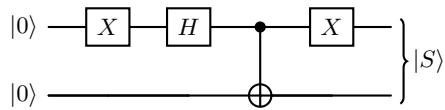

The circuit is straightforward and given in \cref{fig:singlet}.

\section{Generating training data with CHGNET}%
\label{app:data}
As discussed in the main text, we generated our training data for H$_2$O,
CH$_3$O, and CH$_3$CHO using the general force field
CHGNET~\cite{deng2023CHGNetPretrainedUniversal} to calculate an MD trajectory
of a single molecule at $1000$K at $1$fs time steps for a total time of $100$ps.
This data set is made to resemble the rMD17 dataset
\cite{chmiela2017MachineLearningAccurate,christensen2020RoleGradientsMachinea}
in order to explore the potential generalization of our QML potential to the
original dataset. In \cref{tab:md_data} we list some basic properties of the
generated data together with the data for ethanol in the original rMD17
dataset. 
\begin{figure}[tbp]
    \centering
    \begin{quantikz}
        & \gate[2]{RH(\theta_1)} & & \gate[4]{RH(\theta_4)} & \\
        & & \gate[2]{RH(\theta_3)} & \linethrough & \\
        & \gate[2]{RH(\theta_2)} & & \linethrough & \\
        & & & &
    \end{quantikz}
    \caption{Circuit diagram for the 4-qubit register parametrized block RH4.\label{fig:4bitparamblock}}
\end{figure}
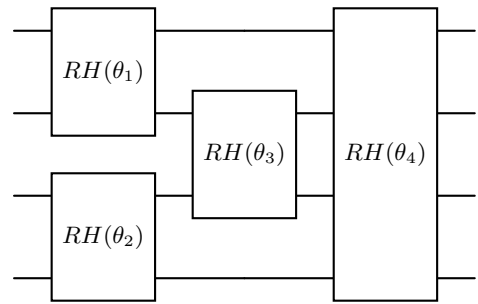
\section{QNN layouts}\label{app:qnn}

In this section, we define the tuple set $I = \{(i_{11}, i_{12}), \ldots, (i_{M1}, i_{M2})\}$, as in \cref{eqn:layer} and \cref{eqn:ang_layer}, specifying the qubit pairs on which the RH gates act (see \cref{tab:qnn_structures}). To simplify notation, we introduce the RH4 gate in \cref{fig:4bitparamblock}, where a quadruple $(i, j, k, l)$ represents an RH4 gate applied to qubits $i, j, k, l$, corresponding to the tuple set $\{(i, j), (k, l), (j, k), (i, l)\}$. 

For the same molecule, we use the same $I$ for both the initial encoding ($N_{\mathrm{q}} = N_a - 1$) and the angular register in the revised encoding. The encoding order of the atoms onto qubits is as follows:

\paragraph{H$_2$O} We take oxygen (O) as the central atom, encoding the two hydrogen (H) atoms' coordinates as inputs to the QNN.

\paragraph{CH$_3$O} We take carbon (C) as the central atom, encoding oxygen (O) on the first qubit, and the remaining three hydrogen (H) atoms on the remaining three qubits.

\paragraph{CH$_3$CHO} The carbon (C) from the CHO group is the central atom, with oxygen (O) on the first qubit, the hydrogen (H) from CHO on the second qubit, the second carbon (C) on the third qubit, and the remaining three hydrogen (H) atoms on the remaining three qubits.

\paragraph{CH$_3$CH$_2$OH (Ethanol)} We use the carbon (C) from the CH$_2$OH group as the central atom, encoding the other carbon (C) on the first qubit, oxygen (O) on the second qubit, the first two hydrogen (H) atoms from CH$_2$OH on the next two qubits, then the three hydrogen (H) atoms from CH$_3$, and finally the hydrogen (H) from OH. This follows the ordering used in the rMD17 dataset.

\bibliography{QMLP}

\end{document}